%% file: main.tex
\documentclass[12pt]{article}

\usepackage[margin=1in]{geometry}
\usepackage{setspace}
\usepackage{amsmath,amssymb,amsfonts,amsthm,bm}
\usepackage{mathtools}
\usepackage{booktabs}
\usepackage{mathrsfs}
\usepackage{graphicx}
\usepackage{float}
\usepackage{multirow}
\usepackage{array}
\usepackage{enumitem}
\usepackage{xcolor}
\usepackage{placeins}
\usepackage{algorithm}
\usepackage{algpseudocode}
\usepackage[round,authoryear]{natbib}
\usepackage{hyperref}

\hypersetup{
  colorlinks=true,
  linkcolor=blue,
  citecolor=blue,
  urlcolor=blue
}



\theoremstyle{plain}
\newtheorem{theorem}{Theorem}[section]

\newtheorem{proposition}[theorem]{Proposition}

\theoremstyle{remark}
\newtheorem{remark}[theorem]{Remark}

\DeclareMathOperator*{\argmin}{\arg\!\min}

\newif\ifblind
\blindfalse

\begin{document}

\begin{center}
{\LARGE \bf Similarity-Based Prediction for Digital Twins: Panel Data, Theory, and Applications}

\vspace{1em}

{\large Ruihang Han$^{a}$ and Li-Hsiang Lin$^{a}$}\\[0.3cm]
{\large $^{a}$ Department of Mathematics and Statistics, Georgia State University, Atlanta, GA 30303}

\end{center}

\vspace{3em}
\thispagestyle{empty}
\begin{abstract}
Prediction from sequential panel data is central to digital-twin modeling, where new panels arrive over time and the predictive system is updated sequentially. Existing methods often rely on temporal proximity, which can fail when similar input–output patterns recur at nonadjacent times or when recent panels differ from the target panel. We propose State-Local Prediction (StaLoP), a nonparametric dynamic panel prediction framework that utilizes information through target-local predictive compatibility. StaLoP represents panels using target-local state vectors, compares historical and target panels via empirical discrepancy scores to determine relevance weights for the target point, and combines these weights with covariate localization. Theoretical results, including bias–variance characterization, asymptotic normality, simultaneous prediction bands, and a target-local GDF-corrected MSPE criterion for panel and model selection, are developed. Extensive simulations validate the performance of StaLoP and support its theoretical properties. Applications to sequence prediction, simulator calibration, variable selection, and county-to-county migration-flow forecasting demonstrate improved out-of-sample prediction and provide scientific insights into the underlying applications.
\end{abstract}


\newpage
\setcounter{page}{1}
\doublespacing

\input{intro}
\input{method}
\input{theory}
\input{extension}

\input{all_simulation}

\input{real_data}

\input{discussion}

\bibliographystyle{plainnat}
\bibliography{ref_intro}

%\appendix
%\input{appendix}

\end{document}

%% file: intro.tex
% ============================================================
% INTRODUCTION -- tightened version with updated canonical references
% ============================================================

\newpage
\section{Introduction}
\label{sec:introduction}

Digital twin modeling has become increasingly important in scientific and
engineering applications where computer and physical experiments are collected
sequentially to study a target phenomenon. Computer experiments provide
surrogate models for predicting responses under new input conditions, and these
predictions guide subsequent physical experiments. The resulting observations
then calibrate the surrogate models, creating a feedback loop that improves
prediction and decision-making over time
\citep{SacksWelchMitchellWynn1989,KennedyOHagan2001}. This feedback perspective
is also emphasized in recent national-level discussions on digital twins, which
highlight the need for statistical theory, uncertainty quantification, and
data-informed decision support \citep{NASEM2024DigitalTwins}. Such sequential feedback naturally produces panel data structure. At each experimental
stage or time point, multiple observations are collected under different input
conditions, forming one panel; over time, these panels describe the evolving
system. In the motivating human-migration example of this study, each time point contains many
origin--destination movement records with economic, environmental, and
conflict-related covariates. These sequential panels can support a digital twin
for forecasting future migration flows under changing conditions and informing
city planning. However, existing prediction strategies often rely on recent
time points, implicitly assuming that temporal proximity implies predictive
relevance. This assumption can fail when an older panel has more similar local input--output behaviour than a recent one, motivating target-specific methods that identify the most relevant historical panels and observations for future
prediction.

This target-specific prediction perspective raises two statistical challenges.
First, the input--output relationship within each panel is often nonlinear and
unknown, and it may change dynamically as the underlying physical system
evolves. Therefore, the target-time mean function should be modeled flexibly as
a nonparametric dynamic function rather than being restricted to a fixed
parametric form. Second, changes in the dynamic mean function need not follow
temporal proximity. Temporal proximity can be useful when the dynamic mean
function evolves smoothly over time, but it can be misleading when similar
patterns recur at nonadjacent times or when recent panels exhibit different
local input--output behavior. These challenges call for a broader digital twin
modeling framework that can flexibly estimate dynamic nonparametric mean
functions while identifying target-relevant historical panels and observations
without relying on time proximity as the default relevance criterion.

These challenges motivate a modeling perspective that differs from existing panel-data approaches. As summarized in Table~\ref{tab:panel_literature_comparison}, representative methods typically do not combine the two features required in the digital twin modeling setting: a nonparametric dynamic mean function and prediction without relying on temporal proximity.  Classical mixed effects models and factor-based panel models mainly address heterogeneity or unobserved common factors, rather than nonparametric prediction of a future
panel \citep{Mundlak1978,HausmanTaylor1981,Pesaran2006,Bai2009}. Time series based panel models introduce temporal dependence through lagged outcomes, but their mean
structures are typically parametric and tied to temporally ordered histories
\citep{ArellanoBond1991,BlundellBond1998}. Nonparametric fixed-effects methods
allow flexible covariate effects, but they do not directly model a
panel-specific dynamic regression surface for future-panel prediction
\citep{SuUllah2006,HendersonCarrollLi2008}. The most related model is time-varying-coefficient modeling,
which allows regression effects to evolve over time or another observed index
\citep{ClevelandDevlin1988,Fan1993,HastieTibshirani1993,
HooverRiceWuYang1998,FanZhang1999,CaiFanLi2000,LinWangWelshCarroll2004,
LiChenGao2011,FengLiPengTong2021}. However, these methods typically utilize
information through temporal or smoothness assumptions, or through structured dependence assumptions, so nearby panels are implicitly treated as more relevant
for prediction.

\begin{table}[ht]
\centering
\caption{Conceptual comparison of representative panel-data methods for
target-future-panel prediction. Here \(V\) denotes yes, \(X\) denotes no, and
N/A denotes not applicable to the primary modeling goal of the corresponding
method class.}
\label{tab:panel_literature_comparison}
\small
\begin{tabular}{
p{0.46\textwidth}
>{\centering\arraybackslash}p{0.23\textwidth}
>{\centering\arraybackslash}p{0.23\textwidth}
}
\hline
Panel data methods
& Target-future-panel nonparametric prediction
& Non-temporal panel relevance weighting \\
\hline

Mixed-effects models

& \(X\)
& N/A \\

Factor-based models

& \(X\)
& N/A \\

Time-series-based panel models

& \(X\)
& \(X\) \\

Nonparametric fixed-effects models

& \(X\)
& N/A \\

Varying-coefficient models

& \(V\)
& \(X\) \\

\textbf{Proposed Method (StaLoP)}
& \(\mathbf{V}\)
& \(\mathbf{V}\) \\

\hline
\end{tabular}
\end{table}

In contrast, we propose \emph{State-Local prediction} (StaLoP). StaLoP allows the panel-specific
mean function to vary nonparametrically across time, while identifying relevant
historical panels through target-local predictive compatibility rather than
temporal proximity. The key idea is to separate two sources of relevance:
observation-level covariate proximity to the target input \citep{fan2018local} and target-local
predictive compatibility with the target time. This separation enables
prediction to use older panels when their local input--output behavior is more
compatible with the target panel than that of more recent panels. Beyond this
methodological contribution, we also develop the theoretical foundation of
StaLoP.  The theories explain why and when target-local borrowing can
improve prediction. Since StaLoP does not assume that recent panels are
automatically most relevant, its performance depends on balancing local
approximation error, transfer bias across panels, and
target-local variance. Our analysis makes this trade-off explicit through a
bias--variance decomposition for the StaLoP Estimator, and then
supports uncertainty quantification through asymptotic normality and
simultaneous prediction bands over a target region. The same trade-off motivates
our model-selection rule: borrowing from more panels may reduce variance but
can introduce less compatible information. We therefore derive a target-local
MSPE decomposition, expressed as residual loss plus a
generalized-degrees-of-freedom correction \citep{Ye1998}, which provides an
interpretable and efficient criterion for selecting tuning parameters of StaLoP. Together, these results establish StaLoP as a novel framework for
nonparametric dynamic panel prediction without relying on temporal proximity as
the default relevance principle.

Although motivated by digital twin panel data for migration studies, the proposed framework can also be applied to sequence data in which only one observation is available at each time point. The method can further be adapted
to computer experiment calibration problems, where a limited physical dataset,
viewed as one target panel, is used to calibrate computer models. These
connections position StaLoP as a statistical surrogate modeling framework with
potential applications in environmental monitoring, smart manufacturing,
infrastructure systems, and other complex physical or cyber--physical systems
\citep{SacksWelchMitchellWynn1989,
KennedyOHagan2001, SantnerWilliamsNotz2018}. In these settings, the proposed
framework aims to improve target-specific prediction by using historical
information according to local predictive relevance. Several applications are demonstrated in the numerical studies.

The remainder of the paper is organized as follows.
Section~\ref{sec:methodology} introduces the proposed methodology, including
the panel-data formulation, target-local state vector construction, empirical
panel discrepancy scores, target relevance weights, and the StaLoP Estimator.
Section~\ref{sec:theory} studies the theoretical properties of the StaLoP Estimator.
Section~\ref{sec:optimal_weight_selection} develops the target-local
GDF-corrected MSPE criterion for selecting the retained-panel count.
Section~\ref{sec:simulation} presents simulation studies evaluating
finite-sample performance.
Section~\ref{sec:application} presents additional applications of State-Local
prediction.
Section~\ref{sec:realdata} presents the empirical migration-flow application.
Section~\ref{sec:discussion} concludes with discussion and possible extensions.

%% file: method.tex
% ============================================================
% METHODOLOGY
% ============================================================
\section{Methodology}
\label{sec:methodology}

We consider prediction in a sequential panel-data setting motivated by
digital-twin feedback systems. At each historical time point \(t=1,\dots,T\),
we observe physical-experiment data
\(\mathcal D_t=\{(\mathbf x_{ti},y_{ti}):i=1,\dots,n_t\}\), where
\(\mathbf x_{ti}\in\mathcal X\subset\mathbb R^d\) is the input condition or
covariate vector, and \(y_{ti}\in\mathbb R\) is the corresponding observed
response. The panel size \(n_t\) may vary across time. We model the
observations as
\begin{equation}
\label{eq:panel_model}
y_{ti}=m_t(\mathbf x_{ti})+\epsilon_{ti},
\qquad
E(\epsilon_{ti}\mid \mathbf x_{ti})=0,
\end{equation}
where \(m_t(\mathbf x)=E(y_{ti}\mid \mathbf x_{ti}=\mathbf x)\) is the
time-specific mean function, and the sequence
\(\{m_t:t=1,\dots,T+1\}\) is not assumed to evolve smoothly in calendar time.
Although \(m_t\) may represent an underlying computer model, we focus on
settings in which this model is expensive to evaluate at all target inputs,
such as when it is a high-fidelity simulator. In such cases, \(m_t\) must be
approximated by a statistical model called surrogate model. Our objective is therefore to
build a surrogate model from the historical panels
\(\mathcal D_1,\dots,\mathcal D_T\) to predict
\(m_{T+1}(\mathbf x^\ast_{T+1})\), without assuming that more temporally recent
panels are necessarily more predictive of the future target.

%As a result, the goal is to predict \(m_{T+1}(\mathbf x^\ast_{T+1})\) at a future target input \(\mathbf x^\ast=\mathbf x^\ast_{T+1}\in\mathcal X\). The sequence \(\{m_t:t=1,\dots,T+1\}\) is not assumed to evolve smoothly in calendar time. Prediction at \(T+1\) therefore requires identifying relevant historical panels according to target-local predictive compatibility, rather than relying on temporal proximity alone, and update the surrogate model.

\subsection{Panel Discrepancy Characterization}
\label{subsec:panel_discrepancy}

The key question is how to measure whether a historical panel is compatible
with the target panel near \(\mathbf x^\ast\). We first show why covariate
proximity alone is insufficient, and then define a target-local discrepancy
measure based on local input--output behavior. For any historical observation
\((\mathbf x_{ti},y_{ti})\), \eqref{eq:panel_model} gives
\begin{equation}
\label{eq:historical_decomposition}
y_{ti}
=
m_{T+1}(\mathbf x^\ast)
+
\{m_{T+1}(\mathbf x_{ti})-m_{T+1}(\mathbf x^\ast)\}
+
\{m_t(\mathbf x_{ti})-m_{T+1}(\mathbf x_{ti})\}
+
\epsilon_{ti}.
\end{equation}
The first bracketed term is local approximation bias \citep{fan2018local}, caused by covariate
mismatch between \(\mathbf x_{ti}\) and \(\mathbf x^\ast\). The second is
transfer bias, caused by mean-function mismatch between panel \(t\) and the
target panel \(T+1\). Thus, borrowing should combine covariate localization to
control local approximation bias with target-local panel compatibility to reduce
transfer bias.

We characterize target-local predictive compatibility through a target-local
basis representation. Let
\(\boldsymbol\phi(\mathbf x,\mathbf x^\ast)\in\mathbb R^L\) denote a local
basis near \(\mathbf x^\ast\). For each \(s=1,\dots,T+1\), write
\begin{equation}
\label{eq:target_basis}
\widetilde m_s(\mathbf x;\mathbf x^\ast)
=
\mathbf b_s(\mathbf x^\ast)^\top
\boldsymbol\phi(\mathbf x,\mathbf x^\ast),
\end{equation}
where \(\mathbf b_s(\mathbf x^\ast)\) is the oracle target-local state vector
of the dynamic mean function at time \(s\). Let \(\mu_{\mathbf x^\ast}\) be a
target-specific measure on \(\mathcal X\), and define
\begin{equation}
\label{eq:basis_measure_matrix}
\mathbf W(\mathbf x^\ast)
=
\int_{\mathcal X}
\boldsymbol\phi(\mathbf x,\mathbf x^\ast)
\boldsymbol\phi(\mathbf x,\mathbf x^\ast)^\top
\,d\mu_{\mathbf x^\ast}(\mathbf x).
\end{equation}
The measure \(\mu_{\mathbf x^\ast}\) determines how local basis directions are
weighted when two panels are compared. In computation,
\(\mathbf W(\mathbf x^\ast)\) may be taken as a kernel-weighted empirical Gram
matrix of the local basis, with ridge stabilization if needed.

\begin{proposition}[Basis-induced representation of panel discrepancy]
\label{thm:basis_panel_discrepancy}
Under \eqref{eq:target_basis} and \eqref{eq:basis_measure_matrix}, for any
historical panel \(t=1,\dots,T\),
\begin{equation}
\label{eq:basis_discrepancy_identity}
\int_{\mathcal X}
\left[
\widetilde m_t(\mathbf x;\mathbf x^\ast)
-
\widetilde m_{T+1}(\mathbf x;\mathbf x^\ast)
\right]^2
d\mu_{\mathbf x^\ast}(\mathbf x)
=
\{
\mathbf b_t(\mathbf x^\ast)
-
\mathbf b_{T+1}(\mathbf x^\ast)
\}^{\top}
\mathbf W(\mathbf x^\ast)
\{
\mathbf b_t(\mathbf x^\ast)
-
\mathbf b_{T+1}(\mathbf x^\ast)
\}.
\end{equation}
If \(\mathbf W(\mathbf x^\ast)\) is positive definite, then the left-hand side
of \eqref{eq:basis_discrepancy_identity} is zero if and only if
\(\mathbf b_t(\mathbf x^\ast)=\mathbf b_{T+1}(\mathbf x^\ast)\).
\end{proposition}

\noindent Proposition~\ref{thm:basis_panel_discrepancy} reduces target-local panel
comparison to a quadratic distance between oracle target-local state vectors,
whose estimators are constructed in the next subsection.

\subsection{Similarity-Based Borrowing for Prediction}
\label{subsec:feasible_similarity_prediction}

For each \(a=1,\dots,T+1\), let \(\mathcal H_a\) be the historical-panel set
used to estimate the target-local state vector for index \(a\). In particular, $
\mathcal H_{T+1}\subset\{1,\dots,T\},$
so that the target-panel responses are not used in constructing the
corresponding target-local state vector. For prediction at the target panel,
all target-local state-vector construction and weight construction are
completed using information available before the target response is evaluated.
Define
\begin{equation}
\label{eq:status_estimation}
\hat{\mathbf z}_a(\mathbf x^\ast)
=
\arg\min_{\mathbf z\in\mathbb R^L}
\sum_{s\in\mathcal H_a}\sum_{i=1}^{n_s}
K_h(\mathbf x_{si}-\mathbf x^\ast)
\left\{
y_{si}
-
\mathbf z^\top
\boldsymbol\phi(\mathbf x_{si},\mathbf x^\ast)
\right\}^2,
\qquad a=1,\dots,T+1.
\end{equation}
Thus, \(\hat{\mathbf z}_a(\mathbf x^\ast)\) summarizes target-local
information near \(\mathbf x^\ast\) using only panels in \(\mathcal H_a\).

Given the estimated target-local state vectors \eqref{eq:status_estimation},
we estimate the panel discrepancy score motivated by
\eqref{eq:basis_discrepancy_identity} as
\begin{equation}
\label{eq:panel_status_score}
\theta_t(\mathbf x^\ast)
=
\{
\hat{\mathbf z}_t(\mathbf x^\ast)
-
\hat{\mathbf z}_{T+1}(\mathbf x^\ast)
\}^{\top}
\widehat{\mathbf W}(\mathbf x^\ast)
\{
\hat{\mathbf z}_t(\mathbf x^\ast)
-
\hat{\mathbf z}_{T+1}(\mathbf x^\ast)
\},
\qquad t=1,\dots,T.
\end{equation}
Here \(\widehat{\mathbf W}(\mathbf x^\ast)\) is the empirical local-basis Gram
matrix
\[
\widehat{\mathbf W}(\mathbf x^\ast)
=
\frac{
\sum_{s=1}^{T}\sum_{i=1}^{n_s}
K_h(\mathbf x_{si}-\mathbf x^\ast)
\boldsymbol\phi(\mathbf x_{si},\mathbf x^\ast)
\boldsymbol\phi(\mathbf x_{si},\mathbf x^\ast)^\top
}{
\sum_{s=1}^{T}\sum_{i=1}^{n_s}
K_h(\mathbf x_{si}-\mathbf x^\ast)
},
\]
with a small ridge term added in numerical implementation if needed. Smaller
\(\theta_t(\mathbf x^\ast)\) indicates stronger target-local predictive
compatibility between historical panel \(t\) and the target panel.

Because the scale of \(\theta_t(\mathbf x^\ast)\) may vary across target
inputs, we can standardize them by utilizing $
\bar\theta(\mathbf x^\ast)=T^{-1}\sum_{t=1}^T\theta_t(\mathbf x^\ast)$ and
$s_\theta(\mathbf x^\ast)=
\left\{
(T-1)^{-1}\sum_{t=1}^T
[\theta_t(\mathbf x^\ast)-\bar\theta(\mathbf x^\ast)]^2
\right\}^{1/2}$,
to get
\begin{equation}
\label{eq:standardized_discrepancy}
\widetilde\theta_t(\mathbf x^\ast)
=
\frac{
\theta_t(\mathbf x^\ast)-\bar\theta(\mathbf x^\ast)
}{
s_\theta(\mathbf x^\ast)+\delta_\theta},
\qquad t=1,\dots,T.
\end{equation}
\noindent The standardized discrepancy score \eqref{eq:standardized_discrepancy} 
can be used to define relevance weights for prediction at the target input $x^{\ast}$, which can be expressed as
\begin{equation}
\label{eq:final_observation_weight}
\hat{w}_{ti}(\mathbf x^\ast)
=
\frac{
\alpha_t(\mathbf x^\ast)K_h(\mathbf x_{ti}-\mathbf x^\ast)
}{
\sum_{s=1}^{T}\sum_{j=1}^{n_s}
\alpha_s(\mathbf x^\ast)K_h(\mathbf x_{sj}-\mathbf x^\ast)
},
\qquad
t=1,\dots,T,\quad i=1,\dots,n_t,
\end{equation}
where $\alpha_t(\mathbf x^\ast)
=
\frac{
\exp\{-\eta\widetilde\theta_t(\mathbf x^\ast)\}
}{
\sum_{s=1}^{T}
\exp\{-\eta\widetilde\theta_s(\mathbf x^\ast)\}
}$ for $t=1, \dots, T$, and \(\eta>0\) controls borrowing concentration across historical panels.

The observation-level prediction weights \eqref{eq:final_observation_weight}
combine panel-level borrowing and covariate localization. With
\eqref{eq:final_observation_weight}, the target-local coefficient vector can be
estimated by
\begin{equation}
\label{eq:weighted_local_prediction_fit}
\hat{\mathbf c}(\mathbf x^\ast)
=
\arg\min_{\mathbf c\in\mathbb R^L}
\sum_{t=1}^{T}\sum_{i=1}^{n_t}
\hat{w}_{ti}(\mathbf x^\ast)
\left\{
y_{ti}
-
\mathbf c^\top
\boldsymbol\phi(\mathbf x_{ti},\mathbf x^\ast)
\right\}^2.
\end{equation}
and the StaLoP Estimate is
\begin{equation}
\label{eq:status_weighted_predictor}
\hat m_{T+1}(\mathbf x^\ast)
=
\hat{\mathbf c}(\mathbf x^\ast)^\top
\boldsymbol\phi(\mathbf x^\ast,\mathbf x^\ast).
\end{equation}
Algorithm~\ref{alg:similarity_local_polynomial} summarizes the aforementioned
procedure for constructing the StaLoP Estimate. Although
\eqref{eq:final_observation_weight} defines weights through target-local
discrepancy, temporal proximity can be viewed as a special case within the
same framework. For example, one may let the panel relevance weight also depend
on a smooth time-decay factor, such as
\(\exp\{-\lambda |t-(T+1)|\}\) with \(\lambda\ge 0\), and then renormalize the
weights. We do not pursue this special case here, since our focus is borrowing
based on target-local predictive compatibility rather than temporal proximity.

\begin{algorithm}[!htbp]
\caption{State-Local prediction (StaLoP) at \(\mathbf x^\ast\)}
\label{alg:similarity_local_polynomial}
\begin{algorithmic}[1]
\Require Historical panels \(\{\mathcal D_t\}_{t=1}^T\), target input
\(\mathbf x^\ast\), historical-panel sets \(\{\mathcal H_a:a=1,\dots,T+1\}\),
basis \(\boldsymbol\phi(\cdot,\mathbf x^\ast)\), kernel \(K_h\), matrix
\(\mathbf W(\mathbf x^\ast)\), small positive constant \(\delta_\theta>0\), and
concentration parameter \(\eta>0\).
\Ensure StaLoP Estimate \(\hat m_{T+1}(\mathbf x^\ast)\).

\State Estimate the target-local state vectors
\(\hat{\mathbf z}_a(\mathbf x^\ast)\), \(a=1,\dots,T+1\), using
\eqref{eq:status_estimation}.

\State Optionally update \(\mathcal H_a\) or preliminary state-estimation
weights during target-local state vector construction; see
Remark~\ref{rem:iterative_state_update} and
Appendix A.

\State Compute the empirical panel discrepancy scores
\(\theta_t(\mathbf x^\ast)\), \(t=1,\dots,T\), using
\eqref{eq:panel_status_score}.

\State Compute the standardized panel discrepancy scores
\(\widetilde\theta_t(\mathbf x^\ast)\), \(t=1,\dots,T\), using
\eqref{eq:standardized_discrepancy}.

\State Construct the observation-level prediction weights
\(w_{ti}(\mathbf x^\ast)\), for \(t=1,\dots,T\) and
\(i=1,\dots,n_t\), using \eqref{eq:final_observation_weight}.

\State Solve \eqref{eq:weighted_local_prediction_fit} and return the StaLoP
Estimate \(\hat m_{T+1}(\mathbf x^\ast)\) from
\eqref{eq:status_weighted_predictor}.
\end{algorithmic}
\end{algorithm}

\begin{remark}[Iterative updates for target-local state vectors and preliminary state-estimation weights]
\label{rem:iterative_state_update}
Starting from preliminary state-estimation weights \(\rho_{as}^{(0)}\), one
may alternate between estimating target-local state vectors and updating
preliminary state-estimation weights. At iteration \(r\), the current weights
\(\rho_{as}^{(r)}\) define a weighted version of
\eqref{eq:status_estimation}. The estimated target-local state vectors are
compared by
$
d_{as}^{(r)}(\mathbf x^\ast)
=
\left\{
\hat{\mathbf z}_a^{(r)}(\mathbf x^\ast)
-
\hat{\mathbf z}_s^{(r)}(\mathbf x^\ast)
\right\}^{\top}
\widehat{\mathbf W}(\mathbf x^\ast)
\left\{
\hat{\mathbf z}_a^{(r)}(\mathbf x^\ast)
-
\hat{\mathbf z}_s^{(r)}(\mathbf x^\ast)
\right\},$
where \(\widehat{\mathbf W}(\mathbf x^\ast)\) is the empirical local-basis Gram
matrix used in \eqref{eq:panel_status_score}.
A generic update is
\[
\rho_{as}^{(r+1)}
=
\frac{
\mathbf 1\{s\in\mathcal H_a\}
\exp\{-\lambda_\rho d_{as}^{(r)}(\mathbf x^\ast)\}
}{
\sum_{\ell\in\mathcal H_a}
\exp\{-\lambda_\rho d_{a\ell}^{(r)}(\mathbf x^\ast)\}
}.
\]
Here \(\lambda_\rho\ge0\) controls concentration; \(\lambda_\rho=0\) gives
uniform weights over \(\mathcal H_a\). This update is optional and affects only
target-local state vector construction. The target relevance weights and
observation-level prediction weights remain defined by
\eqref{eq:final_observation_weight}.
\end{remark}

%% file: theory.tex
% ============================================================
% Section: Theoretical Properties
% ============================================================

\section{Theoretical Properties}
\label{sec:theory}

This section studies the prediction-stage behavior of the StaLoP estimator
constructed in Algorithm~\ref{alg:similarity_local_polynomial}. Throughout this
section, unless otherwise stated, all quantities are evaluated at a fixed target input \(\mathbf x^\ast_{T+1}\) and denoted by \(\mathbf x^\ast_{T+1} \equiv \mathbf x^\ast\) for notational simplicity.

To explore the theoretical property, note that the proposed estimator \eqref{eq:status_weighted_predictor}, which can be expressed as 
\[
\hat m_{T+1}(\mathbf x^\ast)
=
\sum_{t=1}^{T}\sum_{i=1}^{n_t}
\ell_{ti}(\mathbf x^\ast)y_{ti},
\]
where \(\ell_{ti}(\mathbf x^\ast)\) is the equivalent weight induced by the
weighted local linear fit and its detailed expression can be found in equation (25) of
Appendix C. Note that to construct the weight we need the information of
  the
realized historical design denoted by \(\mathscr X_T=\{\mathbf x_{ti}:t=1,\ldots,T,\ i=1,\ldots,n_t\}\), which generates the sigma-field denoted by \(\mathcal F_T(\mathbf x^\ast)\) generated by
\(\mathscr X_T\), and the estimated weights $\{\hat{w}_{ti}(\mathbf x^\ast):t=1,\ldots,T,\ i=1,\ldots,n_t\}$
from \eqref{eq:final_observation_weight}. With these notation, the following theorem makes the bias--variance trade-off of StaLoP explicit.

\begin{theorem}[Conditional bias and variance]
\label{thm:lcdt_bias_variance}
Suppose Conditions B1--B4
in Appendix B hold. Conditional on
\(\mathcal F_T(\mathbf x^\ast)\), there exists a constant \(C>0\) such that
\begin{equation}
\label{eq:bias_bound}
\left|
\mathrm E\!\left[
\hat m_{T+1}(\mathbf x^\ast)-m_{T+1}(\mathbf x^\ast)
\,\middle|\,
\mathcal F_T(\mathbf x^\ast)
\right]
\right|
\le
C
\|\nabla^2 m_{T+1}\|_{\infty,\mathcal B}
h(\mathbf x^\ast)^2
+
B_{\mathrm{tr}}(\mathbf x^\ast),
\end{equation}
where \(\mathcal B\) is a target-local neighborhood of
\(\mathbf x^\ast\) and 
$
B_{\mathrm{tr}}(\mathbf x^\ast)
=
\left|
\sum_{t=1}^{T}\sum_{i=1}^{n_t}
\ell_{ti}(\mathbf x^\ast)
\{m_t(\mathbf x_{ti})-m_{T+1}(\mathbf x_{ti})\}
\right|.
$
\noindent Moreover,
\begin{equation}
\label{eq:variance_bound}
\mathrm{Var}\!\left[
\hat m_{T+1}(\mathbf x^\ast)
\,\middle|\,
\mathcal F_T(\mathbf x^\ast)
\right]
\le
\frac{\sigma^2}{N_{\mathrm{loc}}(\mathbf x^\ast)},
\end{equation}
where $N_{\mathrm{loc}}(\mathbf x^\ast)
=
\left\{
\sum_{t=1}^{T}\sum_{i=1}^{n_t}
\ell_{ti}(\mathbf x^\ast)^2
\right\}^{-1}$.
As a result,
\begin{equation}
\label{eq:mse_bound}
\mathrm E\!\left[
\{\hat m_{T+1}(\mathbf x^\ast)-m_{T+1}(\mathbf x^\ast)\}^2
\,\middle|\,
\mathcal F_T(\mathbf x^\ast)
\right]
\le
2C^2
\|\nabla^2 m_{T+1}\|_{\infty,\mathcal B}^2
h(\mathbf x^\ast)^4
+
2B_{\mathrm{tr}}(\mathbf x^\ast)^2
+
\frac{\sigma^2}{N_{\mathrm{loc}}(\mathbf x^\ast)}.
\end{equation}
\end{theorem}

\noindent
The detailed  proof is given in Appendix C. Theorem~\ref{thm:lcdt_bias_variance} clarifies the statistical trade-off behind
StaLoP borrowing. The estimation error is controlled by three terms: the local approximation bias from smoothing around the target input, the transfer bias from borrowing across panels with different mean functions, and the variance determined by the effective amount of target-local information. This
decomposition explains why borrowing more data is not always beneficial: adding historical observations can reduce variance, but it may also increase transfer bias if those observations come from panels that are not compatible with the
target panel. Importantly, the theorem does not exclude the use of temporal information. Rather, temporal proximity enters the bound only through the realized
observation-level weights. If nearby panels have mean functions close to the target-time mean function, temporal weighting can improve prediction by reducing
\(B_{\mathrm{tr}}(\mathbf x^\ast)\). If recent panels are less compatible with the target panel, however, temporal weighting can increase transfer bias or concentrate the equivalent weights, thereby reducing \(N_{\mathrm{loc}}(\mathbf x^\ast)\). Thus, the theorem provides a unified way to evaluate temporal, non-temporal, or hybrid borrowing rules through the same conditional bias--variance trade-off.

The bias--variance trade-off in
Theorem~\ref{thm:lcdt_bias_variance} further helps the study of the
asymptotic normality of StaLoP. Specifically, the equivalent-weight
representation decomposes the StaLoP prediction error as
\begin{equation}
\label{eq:stalop_error_decomposition}
\begin{aligned}
\hat m_{T_n+1}(\mathbf x^\ast)-m_{T_n+1}(\mathbf x^\ast)
&=
\underbrace{
\sum_{(t,i)\in\mathcal I_n}
\ell_{ti,n}(\mathbf x^\ast)
\{m_{T_n+1}(\mathbf x_{ti,n})-m_{T_n+1}(\mathbf x^\ast)\}
}_{b_{\mathrm{LP},n}(\mathbf x^\ast)}
\\
&\quad+
\underbrace{
\sum_{(t,i)\in\mathcal I_n}
\ell_{ti,n}(\mathbf x^\ast)
\{m_t(\mathbf x_{ti,n})-m_{T_n+1}(\mathbf x_{ti,n})\}
}_{b_{\mathrm{tr},n}(\mathbf x^\ast)}
\\
&\quad+
\underbrace{
\sum_{(t,i)\in\mathcal I_n}
\ell_{ti,n}(\mathbf x^\ast)\epsilon_{ti,n}
}_{\text{weighted random error}},
\end{aligned}
\end{equation}
where \(\mathcal I_n=\{(t,i):t=1,\ldots,T_n,\ i=1,\ldots,n_{t}\}\).
The variance of the weighted random error is denoted by
\[
V_n(\mathbf x^\ast)
=
\mathrm{Var}\!\left[
\sum_{(t,i)\in\mathcal I_n}
\ell_{ti,n}(\mathbf x^\ast)\epsilon_{ti,n}
\,\middle|\,
\mathcal F_n
\right].
\]
Here \(b_{\mathrm{LP},n}(\mathbf x^\ast)\) is the local approximation bias and
\(b_{\mathrm{tr},n}(\mathbf x^\ast)\) is the transfer bias. The next theorem
shows that, after accounting for these two bias components, the StaLoP
prediction error is asymptotically normal.

\begin{theorem}[Asymptotic normality]
\label{thm:state_local_asymptotic_normality}
Under Conditions B1--B5 in
Appendix B. Conditional on \(\mathcal F_n\),
\begin{equation}
\label{eq:clt_biased}
\frac{
\hat m_{T_n+1}(\mathbf x^\ast)
-
m_{T_n+1}(\mathbf x^\ast)
-
b_{\mathrm{LP},n}(\mathbf x^\ast)
-
b_{\mathrm{tr},n}(\mathbf x^\ast)
}{
V_n(\mathbf x^\ast)^{1/2}
}
\Rightarrow N(0,1).
\end{equation}
\end{theorem}

%If \begin{equation} \label{eq:bias_negligible} b_{\mathrm{LP},n}(\mathbf x^\ast)+b_{\mathrm{tr},n}(\mathbf x^\ast)=o\{V_n(\mathbf x^\ast)^{1/2}\},\end{equation}then\begin{equation}\label{eq:clt_centered}\frac{\hat m_{T_n+1}(\mathbf x^\ast)-m_{T_n+1}(\mathbf x^\ast)}{V_n(\mathbf x^\ast)^{1/2}}\Rightarrow N(0,1).\end{equation}

\noindent Theorem~\ref{thm:state_local_asymptotic_normality} shows that, after subtracting both the local approximation bias
and the transfer bias, the standardized error converges in distribution
to \(N(0,1)\). The detailed proof is given in
Appendix D.

The pointwise normal approximation in
Theorem~\ref{thm:state_local_asymptotic_normality} shows that the StaLoP
prediction error is asymptotically normal after accounting for the local
approximation and transfer-bias components. This motivates using a
bias-corrected StaLoP center when constructing uncertainty intervals,
especially when simultaneous coverage is required over a target region
\(\mathcal X_0\). We therefore consider a simultaneous band for the target mean
function based on the bias-corrected StaLoP center
\(\hat m^{\mathrm{bc}}_{T_n+1}(\mathbf x)\) and the feasible standard error
\(\widehat{\mathrm{SE}}_{n}(\mathbf x)\). The construction of
\(\hat m^{\mathrm{bc}}_{T_n+1}(\mathbf x)\), its equivalent weights, the
corresponding residual bias, and the oracle conditional variance are given in
Appendix E. Let
\(c_{1-\alpha}\) denote the \((1-\alpha)\)-quantile of the limiting
distribution of
\(\sup_{\mathbf x\in\mathcal X_0}|G(\mathbf x)|\), taken at a continuity
point, where \(G\) is the limiting standardized Gaussian process. The following
theorem formalizes this idea by showing that the standardized bias-corrected
StaLoP process admits a uniform Gaussian approximation over \(\mathcal X_0\),
which yields asymptotically valid simultaneous coverage.

\begin{theorem}[Simultaneous band for the target mean function]
\label{thm:simultaneous_critical_constant}
Suppose Conditions B6--B13
in Appendix B hold. Then
\begin{equation}
\label{eq:simultaneous_sup_bound}
P\left[
\sup_{\mathbf x\in\mathcal X_0}
\left|
\frac{
\hat m^{\mathrm{bc}}_{T_n+1}(\mathbf x)-m_{T_n+1}(\mathbf x)
}{
\widehat{\mathrm{SE}}_{n}(\mathbf x)
}
\right|
\le
c_{1-\alpha}
\right]
\longrightarrow
1-\alpha .
\end{equation}
Consequently, the simultaneous band for the target mean function
\[
\mathcal C_{1-\alpha,n}(\mathbf x)
=
\left[
\hat m^{\mathrm{bc}}_{T_n+1}(\mathbf x)
-
c_{1-\alpha}\widehat{\mathrm{SE}}_{n}(\mathbf x),
\ 
\hat m^{\mathrm{bc}}_{T_n+1}(\mathbf x)
+
c_{1-\alpha}\widehat{\mathrm{SE}}_{n}(\mathbf x)
\right],
\qquad
\mathbf x\in\mathcal X_0,
\]
satisfies
\begin{equation}
\label{eq:simultaneous_coverage}
P\left[
m_{T_n+1}(\mathbf x)\in \mathcal C_{1-\alpha,n}(\mathbf x)
\text{ for all }\mathbf x\in\mathcal X_0
\right]
\longrightarrow
1-\alpha.
\end{equation}
as $n$ goes to $\infty$.
\end{theorem}

\noindent The proof of
Theorem~\ref{thm:simultaneous_critical_constant} is given in
Appendix E. Theorem~\ref{thm:simultaneous_critical_constant} extends pointwise uncertainty
quantification for the target mean function to a uniform statement over
\(\mathcal X_0\) for the realized-weight bias-corrected StaLoP procedure.
In computation, \(c_{1-\alpha}\) is approximated on a dense grid using the
covariance matrix induced by the bias-corrected equivalent weights. Implementation details of the simultaneous prediction intervals are given in
Appendix F. 

%% file: extension.tex
% ============================================================
% Section 4: Optimal Retained-Panel Selection
% ============================================================

\section{Optimal Retained-Panel Selection}
\label{sec:optimal_weight_selection}

The results in Section~\ref{sec:theory} show that the StaLoP Estimate depends
on a bias--variance trade-off induced by the observation-level prediction
weights. Borrowing from fewer historical panels may reduce transfer bias, but
it may also reduce the effective sample size. Borrowing from more historical
panels may reduce variance, but it may include panels with less predictive
compatibility. Thus, selecting which panels for use is important. We proposed selecting the retained-panel count at each target
input by minimizing an estimated mean-function MSPE criterion.

For each candidate retained-panel count \(K\in\mathcal K\), the StaLoP
from \eqref{eq:status_weighted_predictor} can be written as
\[
\hat{\mathbf y}_K
=
\mathbf H_K(\mathbf x^\ast)\mathbf y,
\]
where \(\mathbf H_K(\mathbf x^\ast)=\mathbf\Phi(\mathbf x^\ast)
\{\mathbf\Phi(\mathbf x^\ast)^\top\bar{\mathbf D}_K(\mathbf x^\ast)
\mathbf\Phi(\mathbf x^\ast)\}^{-1}
\mathbf\Phi(\mathbf x^\ast)^\top\bar{\mathbf D}_K(\mathbf x^\ast)\) is the
weighted hat matrix under retained count \(K\), and
\(\bar{\mathbf D}_K(\mathbf x^\ast)\) be the diagonal matrix of the normalized
candidate observation-level weights \(\hat{w}_{ti,K}(\mathbf x^\ast)\) from \eqref{eq:status_weighted_predictor}, ordered
consistently with the stacked response vector \(\mathbf y\). Thus,
\(\operatorname{tr}\{\bar{\mathbf D}_K(\mathbf x^\ast)\}=1\). In the following discussion,  denote the weighted norm
$
\|\mathbf v\|_{\bar{\mathbf D}_K(\mathbf x^\ast)}^2
=
\mathbf v^\top
\bar{\mathbf D}_K(\mathbf x^\ast)
\mathbf v.$
The matrix forms of \(\mathbf H_K(\mathbf x^\ast)\) and
\(\bar{\mathbf D}_K(\mathbf x^\ast)\) are given in
Appendix H.

We now define the population target used to compare different candidate
retained-panel counts. Following the conditional viewpoint in
Section~\ref{sec:theory}, let \(\mathcal F_T(\mathbf x^\ast)\) denote the
sigma-field generated by the realized historical design \(\mathscr X_T\). Let
\(\boldsymbol\mu=E(\mathbf y\mid\mathscr X_T)\) be the conditional mean vector
of the stacked historical responses, and let
\(\mathbf y^\circ=\boldsymbol\mu+\boldsymbol\epsilon^\circ\) be an independent
replicate response vector satisfying
\(E^\circ\{\boldsymbol\epsilon^\circ\mid\mathcal F_T(\mathbf x^\ast)\}=\mathbf 0\)
and
\(\mathrm{Var}^\circ\{\boldsymbol\epsilon^\circ\mid
\mathcal F_T(\mathbf x^\ast)\}=\sigma^2\mathbf I\). For each candidate
retained-panel count \(K\in\mathcal K\), we define the mean squared prediction
error (MSPE) of StaLoP by
\begin{equation}
\label{eq:true_target_local_mspe}
\mathrm{MSPE}_{0}(K;\mathbf x^\ast)
=
E^\circ\left[
\left\|
\mathbf H_K(\mathbf x^\ast)\mathbf y^\circ-\boldsymbol\mu
\right\|_{\bar{\mathbf D}_K(\mathbf x^\ast)}^2
\,\middle|\,
\mathcal F_T(\mathbf x^\ast)
\right].
\end{equation}
This criterion measures estimation error for the conditional mean and excludes
the irreducible noise variance of a new response. If
\(\mathrm{MSPE}_{0}(K;\mathbf x^\ast)\) were known, one could select
\(K_0^\ast=\argmin_{K\in\mathcal K}\mathrm{MSPE}_{0}(K;\mathbf x^\ast)\). The
empirical criterion below provides a plug-in approximation to this population
target.

\begin{theorem}[Weighted MSPE identity]
\label{thm:fixed_weight_mspe_identity}
Suppose Conditions G1--G3 hold. Then, for each fixed
\(K\in\mathcal K\),
\begin{equation}
\label{eq:fixed_weight_mspe_identity}
\mathrm{MSPE}_{0}(K;\mathbf x^\ast)
=
E^\circ\{R_K^\circ(\mathbf x^\ast)\mid\mathcal A_K(\mathbf x^\ast)\}
+
2\sigma^2\mathrm{GDF}_K(\mathbf x^\ast)
-
\sigma^2,
\end{equation}
where \(R_K^\circ(\mathbf x^\ast)=\|\mathbf y^\circ-\mathbf H_K(\mathbf
x^\ast)\mathbf y^\circ\|_{\bar{\mathbf D}_K(\mathbf x^\ast)}^2\) is the
weighted residual loss and
\(\mathrm{GDF}_K(\mathbf x^\ast)=\operatorname{tr}\{\bar{\mathbf D}_K(\mathbf
x^\ast)\mathbf H_K(\mathbf x^\ast)\}\) is the weighted generalized degrees
of freedom. If the normalized-loss condition is not imposed, the final term is
\(-\sigma^2\operatorname{tr}\{\bar{\mathbf D}_K(\mathbf x^\ast)\}\).
\end{theorem}

\noindent The proof is given in Appendix H.
Theorem~\ref{thm:fixed_weight_mspe_identity} expresses MSPE function \ref{eq:true_target_local_mspe}
as an expected replicate residual loss plus a weighted optimism correction.
Since the expected residual loss and \(\sigma^2\) are unknown, we replace them
by empirical counterparts. For the observed response vector \(\mathbf y\),
define \(R_K(\mathbf x^\ast)=\|\mathbf y-\mathbf H_K(\mathbf x^\ast)\mathbf y
\|_{\bar{\mathbf D}_K(\mathbf x^\ast)}^2\). With a variance estimate
\(\widehat\sigma^2\), define the estimated MSPE criterion by
\begin{equation}
\label{eq:estimated_mspe_criterion}
\widehat{\mathrm{MSPE}}(K;\mathbf x^\ast)
=
R_K(\mathbf x^\ast)
+
2\widehat\sigma^2\mathrm{GDF}_K(\mathbf x^\ast)
-
\widehat\sigma^2,
\qquad
\widehat K
=
\argmin_{K\in\mathcal K}
\widehat{\mathrm{MSPE}}(K;\mathbf x^\ast).
\end{equation}
The variance estimate \(\widehat\sigma^2\) may be obtained from residuals of a
preliminary local regression fit or from a pooled residual variance estimator.
The estimated criterion in \eqref{eq:estimated_mspe_criterion} is a plug-in
analogue of \eqref{eq:true_target_local_mspe}: it replaces the conditional
expected replicate residual loss by its observed counterpart and replaces
\(\sigma^2\) by \(\widehat\sigma^2\). Its role is to select a retained-panel
count that balances weighted residual fit and smoother complexity, not to
estimate the oracle transfer bias directly. The next result justifies using \eqref{eq:estimated_mspe_criterion} to select
the retained-panel count over \(\mathcal K\).

\begin{theorem}[Consistency of the estimated MSPE criterion]
\label{thm:estimated_mspe_consistency}
Suppose Conditions G1--G6 hold. Then
\begin{equation}
\label{eq:uniform_mspe_consistency}
\sup_{K\in\mathcal K}
\left|
\widehat{\mathrm{MSPE}}(K;\mathbf x^\ast)
-
\mathrm{MSPE}_{0}(K;\mathbf x^\ast)
\right|
=
o_p(1).
\end{equation}
Consequently, \(\mathrm{MSPE}_{0}(\widehat K;\mathbf x^\ast)-\min_{K\in
\mathcal K}\mathrm{MSPE}_{0}(K;\mathbf x^\ast)=o_p(1)\). If
Condition G7 also holds, then
\(P(\widehat K=K_0^\ast)\to1\).
\end{theorem}

\noindent The detailed proof is given in
Appendix I.
Theorem~\ref{thm:estimated_mspe_consistency} shows that the estimated MSPE
criterion uniformly approximates the population target
\(\mathrm{MSPE}_{0}(K;\mathbf x^\ast)\) over the candidate set
\(\mathcal K\). Based
on \eqref{eq:estimated_mspe_criterion}, retained-panel selection is implemented
by evaluating \(\widehat{\mathrm{MSPE}}(K;\mathbf x^\ast)\) over \(K\in
\mathcal K\), selecting the minimizer \(\widehat K\), and using the selected
retained-panel count to construct the final weighted local-linear StaLoP
Estimator. The procedure is summarized in Algorithm \ref{alg:optimal_weight_selection}.

\begin{algorithm}[!htbp]
\caption{MSPE tuning for retained-panel selection}
\label{alg:optimal_weight_selection}
\begin{algorithmic}[1]
\Require Historical panels \(\{\mathcal D_t\}_{t=1}^T\), target input
\(\mathbf x^\ast\), candidate retained-panel counts \(\mathcal K\), empirical
panel discrepancy scores \(\{\theta_t(\mathbf x^\ast)\}_{t=1}^T\), selected
kernel \(K_h\), selected concentration parameter \(\eta\), and variance
estimate \(\widehat\sigma^2\).
\Ensure Selected retained-panel count \(\widehat K\) and StaLoP Estimate
\(\widehat m_{T+1}(\mathbf x^\ast)\) computed from the selected weighted
local-linear fit.

\State Rank historical panels by the empirical panel discrepancy scores
\(\theta_t(\mathbf x^\ast)\).

\For{each \(K\in\mathcal K\)}
    \State Retain the \(K\) panels with the smallest empirical panel
    discrepancy scores.
    \State Construct the normalized observation-level weights
    \(\{w_{ti,K}(\mathbf x^\ast)\}\), the diagonal weighting matrix
    \(\bar{\mathbf D}_K(\mathbf x^\ast)\), and the candidate retained-panel
    smoother \(\widehat{\mathbf y}_K=\mathbf H_K(\mathbf x^\ast)\mathbf y\).
    \State Compute \(R_K(\mathbf x^\ast)\), \(\mathrm{GDF}_K(\mathbf x^\ast)\),
    and \(\widehat{\mathrm{MSPE}}(K;\mathbf x^\ast)\) using
    \eqref{eq:estimated_mspe_criterion}.
\EndFor

\State Select \(\widehat K=\argmin_{K\in\mathcal K}
\widehat{\mathrm{MSPE}}(K;\mathbf x^\ast)\).

\State Return \(\widehat K\) and the StaLoP Estimate
\(\widehat m_{T+1}(\mathbf x^\ast)\) computed from the selected weighted
local-linear fit.
\end{algorithmic}
\end{algorithm}

\begin{remark}[Panel-level selection and within-panel localization]
\label{rem:panel_selection_within_panel_localization}
The retained-panel selection step operates at the panel level because the
empirical panel discrepancy score is designed to measure target-local
predictive compatibility between dynamic mean functions. Within a retained
panel, observation-level relevance is handled continuously through the kernel
factor \(K_h(\mathbf x_{ti}-\mathbf x^\ast)\) and the weighted local linear
fit. Introducing a second hard selection step within each retained panel would
add another discrete tuning parameter and may reduce the target-local
effective sample size, while providing limited benefit when covariate
localization already downweights observations far from \(\mathbf x^\ast\).
Such within-panel screening may be useful in settings with known subpanel
heterogeneity, but it is not part of the default StaLoP construction studied
here.
\end{remark}

%% file: all_simulation.tex
% ============================================================
% Section: Simulation Studies
% ============================================================

\section{Simulation Studies}
\label{sec:simulation}

We conduct dynamic panel simulation studies to evaluate whether the proposed StaLoP estimator can identify historical panels with target-local predictive compatibility. The simulation is designed so that similar mean-function patterns may recur at nonadjacent time points, making temporal proximity alone insufficient for prediction. This setting is analogous to the migration study in Section 7, where the most recent panels may not be the most informative for prediction, whereas panels corresponding to previous flood-event periods may provide stronger target-local predictive compatibility. The goal is to validate the proposed target-local discrepancy construction and to examine whether the empirical results support the theoretical bias--variance, normal approximation, simultaneous-band, and MSPE-based retained-panel selection results.

We generate a balanced dynamic panel with $N$ units over $T$ time points after discarding a burn-in period of 50 observations. Each unit has a two-dimensional unit-specific mean $\boldsymbol\mu_i\sim N(0,I_2)$. Conditional on $\boldsymbol\mu_i$, the covariates follow unit-specific mean-reverting AR(1) processes with autoregressive coefficient $\rho_x=0.6$, marginal innovation variance one, and innovation correlation $\rho_\eta=0.5$. Time-specific heterogeneity is generated through a common latent class $S_t\in\{1,\ldots,4\}$, evolving according to a cyclic transition rule with switching probability $0.35$. The response is generated from
\[
y_{it}=g_{S_t}(x_{i,t}^{(1)},x_{i,t}^{(2)})+a_i+\varepsilon_{it},
\]
where $a_i\sim N(0,0.5^2)$, $\varepsilon_{it}\sim N(0,1)$, and $g_s$ is a class-specific nonlinear mean function. We consider $(T,N)\in\{(30,40),(30,200),(120,40),(120,200)\}$. In each replication, the target time is $t^\ast=\lfloor 0.7T\rfloor$, and all methods use only observations from $t<t^\ast$. Prediction is evaluated for all $N$ units at $t^\ast$, with performance measured against the noise-free target mean $m_{i,t^\ast}=g_{S_{t^\ast}}(x_{i,t^\ast}^{(1)},x_{i,t^\ast}^{(2)})+a_i$. The reported MSPE is $N^{-1}\sum_{i=1}^N(\hat m_{i,t^\ast}-m_{i,t^\ast})^2$, averaged over $R=300$ Monte Carlo replications. Thus, performance is evaluated conditional on the realized unit effects and
measures recovery of the conditional mean rather than independent observation
noise.

StaLoP constructs target-local state vectors, computes empirical panel discrepancy scores, and combines panel-level relevance with covariate localization in a weighted local-linear fit. The retained-panel count is selected at each target input using the GDF-corrected MSPE criterion from Section~\ref{sec:optimal_weight_selection}. Other tuning parameters, including those used for the optimal retained panel and for computing $c_{1-\alpha}$ in the simultaneous-band procedure discussed in Appendix~H, are also tuned through the proposed MSPE approach. We compare StaLoP with three baselines. \emph{Global Linear} pools all pre-target observations in a linear regression and therefore ignores nonlinear dynamic structure and recurrent panel patterns. \emph{Time-Local} uses only the most recent $L$ pre-target periods, with $L$ selected by pre-target validation, and therefore represents borrowing based on temporal proximity. \emph{KNN-Covariate} uses a local linear regression based on the $k$ nearest historical observations in standardized covariate space, with $k$ selected by pre-target validation; this method uses covariate localization but does not evaluate target-local mean-function compatibility.

\begin{table}[t]
\centering
\caption{\textbf{Dynamic panel simulation: MSPE and runtime.} Entries are Monte Carlo mean MSPE or runtime with empirical standard deviation in parentheses, over $R=300$ replications. Runtime is the average wall-clock time per replication in seconds.}
\label{tab:dynamic_panel_mspe_runtime_draft}
\small
\setlength{\tabcolsep}{4.5pt}
\begin{tabular}{cc l c c}
\toprule
$T$ & $N$ & Method & MSPE (SD) & Runtime (SD) \\
\midrule
30 & 40
& Global Linear          & 17.129 (14.325) & 0.0033 (0.00048) \\
   &
& Time-Local             & 17.076 (13.821) & 0.0036 (0.00052) \\
   &
& KNN-Covariate          &  5.403 (6.132)  & 0.0062 (0.00107) \\
   &
& StaLoP                 &  \textbf{3.956} (3.309)  & 0.1227 (0.00865) \\
\midrule
30 & 200
& Global Linear          & 17.137 (9.104) & 0.0255 (0.00330) \\
   &
& Time-Local             & 17.131 (9.114) & 0.0243 (0.00361) \\
   &
& KNN-Covariate          &  5.075 (3.018) & 0.0407 (0.00534) \\
   &
& StaLoP                 &  \textbf{3.442} (2.561) & 0.7485 (0.04913) \\
\midrule
120 & 40
& Global Linear          & 16.694 (12.246) & 0.0049 (0.00028) \\
    &
& Time-Local             & 16.851 (12.039) & 0.0039 (0.00029) \\
    &
& KNN-Covariate          &  4.748 (4.539)  & 0.0082 (0.00052) \\
    &
& StaLoP                 &  \textbf{3.053} (2.184)  & 0.3378 (0.02021) \\
\midrule
120 & 200
& Global Linear          & 17.372 (9.455) & 0.0696 (0.00509) \\
    &
& Time-Local             & 17.417 (9.453) & 0.0359 (0.00346) \\
    &
& KNN-Covariate          &  5.026 (3.415) & 0.0883 (0.00638) \\
    &
& StaLoP                 &  \textbf{3.021} (1.922) & 2.1453 (0.34638) \\
\bottomrule
\end{tabular}
\end{table}

Table~\ref{tab:dynamic_panel_mspe_runtime_draft} first supports the panel-discrepancy motivation. Global Linear and Time-Local both have MSPEs close to 17 in all four settings, indicating that global pooling and temporal proximity do not resolve the recurrent heterogeneous mean-function structure. KNN-Covariate reduces MSPE to about 5, showing that covariate localization is useful. StaLoP further reduces MSPE in every setting, which supports the proposed basis-induced panel discrepancy idea: covariate closeness alone is not sufficient, and borrowing improves when historical panels are selected by target-local predictive compatibility.

The table also supports Theorem~\ref{thm:lcdt_bias_variance} on conditional bias and variance. The theorem decomposes StaLoP error into local approximation bias, transfer bias, and a variance term governed by target-local effective information. The comparison between KNN-Covariate and StaLoP shows the role of transfer-bias control: both methods localize in covariate space, but StaLoP also downweights panels whose target-local mean behavior is less compatible with the target panel. The decrease in StaLoP MSPE when $T$ increases from 30 to 120 also agrees with the theorem: additional historical panels help when the method can retain compatible panels rather than pooling all panels indiscriminately.

The runtime column is most directly related to the Section~\ref{sec:optimal_weight_selection} optimal retained-panel procedure. StaLoP is computationally slower because it estimates target-local state vectors, ranks panels by empirical discrepancy, evaluates candidate retained-panel counts, and solves weighted local fits. The runtime increases from $0.1227$ seconds at $(T,N)=(30,40)$ to $2.1453$ seconds at $(T,N)=(120,200)$. This confirms the computational cost of the MSPE-based retained-panel selection, but the cost remains moderate in these simulation sizes. The results therefore support the practical implementation of the weighted MSPE identity in Theorem~\ref{thm:fixed_weight_mspe_identity} and the consistency-motivated retained-panel selection rule in Theorem~\ref{thm:estimated_mspe_consistency}.

\begin{table}[t]
\centering
\caption{Simultaneous prediction band performance under the dynamic panel design.
The nominal simultaneous coverage level is $1-\alpha=0.9$. SimCov reports the empirical simultaneous coverage probability of the proposed method, and AvgWidth reports the average band width.}
\label{tab:simultaneous_band_performance_draft}
\small
\begin{tabular}{cccc}
\toprule
$T$ & $N$ & SimCov  & AvgWidth \\
\midrule
30  & 40  & 0.943 & 22.77 \\
30  & 200 & 0.910 & 16.18 \\
120 & 40  & 0.947 & 17.70 \\
120 & 200 & 0.920 & 14.64 \\
\bottomrule
\end{tabular}
\end{table}

Table~\ref{tab:simultaneous_band_performance_draft} evaluates the simultaneous-band result in Theorem~\ref{thm:simultaneous_critical_constant}. The detailed implementation of the band is provided in Appendix F. Across all four panel settings, the empirical simultaneous coverage is above the nominal level $1-\alpha=0.9$. Moreover, for each fixed value of $T$, increasing $N$ moves the empirical coverage closer to the nominal level while reducing the average band width. This pattern is consistent with the theory: a larger within-panel sample provides more target-local information, improves the stability of the equivalent-weight covariance approximation, and yields less conservative simultaneous bands. Therefore, the results mainly support Theorem~\ref{thm:simultaneous_critical_constant}, which establishes asymptotically valid simultaneous coverage for the target mean function. They also provide indirect support for Theorem~\ref{thm:state_local_asymptotic_normality}, because the simultaneous-band construction relies on the bias-corrected Gaussian approximation to the StaLoP error process.

The average band width decreases when $N$ increases, from 22.77 to 16.18 for $T=30$ and from 17.70 to 14.64 for $T=120$. This pattern is consistent with the variance component in Theorem~\ref{thm:lcdt_bias_variance}: larger within-panel sample size increases target-local information and reduces uncertainty. The larger selected calibration factors for $T=120$ reflect the additional adaptation in longer histories, including state construction, panel ranking, and retained-panel selection before forming the simultaneous band. Thus, Table~\ref{tab:simultaneous_band_performance_draft} confirms that the finite-sample simultaneous-band procedure is aligned with the theoretical approximation and coverage result.

\section{Applications of State-Local Prediction}
\label{sec:application}

We present three applications to illustrate how the proposed StaLoP framework
can be used beyond the dynamic panel simulation. The examples cover
sequence data prediction, simulator calibration, and target-relevant variable
selection. In each case, the goal is to examine whether StaLoP improves prediction and gain more scientific insights.

\subsection{Sequence Data}
\label{sec:application_sequence}

We first consider rolling one-step prediction for a smooth periodic signal
observed with additive Gaussian noise. The signal is
\(f(t)=A\sin(2\pi t)\), with \(A=0.75\), and observations follow
\(y(t)=f(t)+\epsilon(t)\), where
\(\epsilon(t)\stackrel{\mathrm{i.i.d.}}{\sim}N(0,\sigma^2)\) and
\(\sigma=0.1\). Data are generated on the grid
\(t\in\{0,0.01,0.02,\ldots,1.60\}\). For each rolling target time
\(t_a\ge1.10\), the prediction target is the noise-free one-step-ahead mean
\(f(t_{a+1})\), where \(t_{a+1}=t_a+0.01\). All methods use only observations
available before the target time. For each grid time \(t_s\), a one-sided
target-local state vector is constructed from the most recent
\(K_{\mathrm{state}}=7\) observations using a local linear intercept--slope
summary \(Z_s=(\hat b_{0,s},\hat b_{1,s})^\top\). The state-construction window
\(K_{\mathrm{state}}\) is fixed across Monte Carlo replications in this
application. The comparison includes Global Linear, Time-Local, KNN-Covariate,
and StaLoP. For StaLoP, the concentration parameter is selected by pre-target
rolling validation over a finite grid, and the retained-panel count is selected
by the target-local GDF-corrected MSPE criterion. This application evaluates
whether StaLoP can borrow from historical states with stronger target-local
predictive compatibility, even when they are not the most recent observations.

\begin{table}[t]
\centering
\caption{\textbf{Sequence-data application: rolling one-step prediction.}
Entries are Monte Carlo mean MSPE with empirical standard deviation in
parentheses across \(R=300\) replications. The overall entry averages squared
prediction errors over all rolling one-step target times within each
replication before averaging across replications.}
\label{tab:sin_mspe_compact}
\begin{tabular}{lc}
\toprule
Method & Overall MSPE (SD) \\
\midrule
Global Linear              & 0.005804 (0.001593) \\
Time-Local                 & 0.010292 (0.005733) \\
KNN-Covariate              & 0.008781 (0.002997) \\
StaLoP                     & \textbf{0.005580 (0.002304)} \\
\bottomrule
\end{tabular}
\end{table}
\FloatBarrier

Table~\ref{tab:sin_mspe_compact} shows that StaLoP has the lowest overall
MSPE. The gain over Global Linear is small because the signal is smooth and a
simple linear predictor is already effective for one-step prediction. The
improvement is clearer relative to Time-Local and KNN-Covariate. Time-Local
uses recency as the borrowing rule, while KNN-Covariate uses only covariate
localization. StaLoP improves on both by selecting observations according to
target-local state compatibility. Averaged over Monte Carlo replications and rolling target times, StaLoP
selected a mean retained-panel count of approximately \(36\) and a mean
concentration parameter of approximately \(0.56\), indicating moderate
borrowing from compatible historical states.

\subsection{Simulator Calibration}
\label{sec:application_computer_experiment}

The second application examines simulator calibration when the calibrated
simulator differs from the oracle data-generating mechanism. The input is
\(X=(X_1,X_2)^\top\), and the oracle conditional mean is generated from a
constant elasticity of substitution (CES) function,
\[
f_{\mathrm{CES}}(X_1,X_2;\theta)
=
A\{\delta X_1^\rho+(1-\delta)X_2^\rho\}^{1/\rho},
\]
with \(\theta_{\mathrm{true}}=(A,\delta,\rho)=(1.0,0.6,-0.5)\). In each
replication, a real sample of size \(n_{\mathrm{real}}=300\) is generated with
\[
(\log X_1,\log X_2)^\top \sim N(0,\Sigma),
\]
where \(\Sigma_{11}=\Sigma_{22}=0.5\) and
\(\Sigma_{12}=\Sigma_{21}=0.2\). The observed response includes
multiplicative log-normal noise. A calibrated parameter \(\hat\theta\) is
obtained by constrained least squares with \(\rho\in[-0.95,0.95]\). Given
\(\hat\theta\), a simulator dataset of size \(n_{\mathrm{sim}}=500\) is
generated over the central quantile range of the real sample. Evaluation is
against the oracle mean \(f_{\mathrm{CES}}(\cdot;\theta_{\mathrm{true}})\), so
the experiment isolates the effect of calibration mismatch. StaLoP constructs
target-local state vectors from nearest-neighbor neighborhoods in standardized
input space and ranks simulator points by empirical panel discrepancy score
from the target input. The concentration parameter \(\eta\) and the
retained-panel count \(q\) are selected over finite grids by the target-local
GDF-corrected MSPE criterion.

\begin{table}[t]
\centering
\caption{\textbf{Simulator calibration: predictive performance under
increasing calibration noise.} Entries are Monte Carlo mean MSPE with empirical
standard deviation in parentheses across \(R=300\) replications. For StaLoP,
the concentration parameter \(\eta\) and the retained-panel count are selected
over finite grids by the target-local GDF-corrected MSPE criterion.}
\label{tab:ces_mspe}
\begin{tabular}{llc}
\toprule
\(\sigma_\varepsilon\) & Method & MSPE (SD) \\
\midrule
0.1 & Global Linear              & 0.040697 (0.008115) \\
    & KNN-Covariate              & 0.022323 (0.005428) \\
    & StaLoP                     & \textbf{0.001137 (0.000575)} \\
\midrule
0.2 & Global Linear              & 0.042778 (0.009170) \\
    & KNN-Covariate              & 0.025696 (0.007459) \\
    & StaLoP                     & \textbf{0.002376 (0.002082)} \\
\midrule
0.3 & Global Linear              & 0.047960 (0.012376) \\
    & KNN-Covariate              & 0.033151 (0.012721) \\
    & StaLoP                     & \textbf{0.006200 (0.006431)} \\
\bottomrule
\end{tabular}
\end{table}
\FloatBarrier

Table~\ref{tab:ces_mspe} shows that StaLoP has the lowest mean MSPE at each
calibration-noise level. As \(\sigma_\varepsilon\) increases, the MSPE of all
methods increases, reflecting the larger calibration error. Global Linear is
limited by the nonlinear CES mean function, and KNN-Covariate improves on it by
using local input information. The resulting StaLoP procedure jointly combines target-local compatibility
borrowing with GDF-corrected selection of \(\eta\) and the retained-panel count
over finite grids.

\subsection{Target-Relevant Variable Selection}
\label{sec:application_variable_selection}

The third application examines whether the target-local GDF-corrected MSPE
criterion can select target-relevant variables when irrelevant variables are
present. The data-generating process follows the dynamic panel design described
above, where the true dynamic mean function depends only on
\(x_{i,t}^{(1)}\) and \(x_{i,t}^{(2)}\). We augment the input vector with two
independent noise variables, \(z_{i,t}^{(1)}\) and \(z_{i,t}^{(2)}\), which do
not enter the data-generating mechanism. The full candidate vector is
\[
\mathbf w_{i,t}
=
(x_{i,t}^{(1)},x_{i,t}^{(2)},z_{i,t}^{(1)},z_{i,t}^{(2)})^\top,
\]
giving \(2^4=16\) candidate subsets. For each subset, the entire StaLoP
procedure is recomputed using only the selected variables, including covariate
distances, local bases, target-local state vectors, empirical panel discrepancy
scores, target relevance weights, and StaLoP estimates. The selected model is
the subset minimizing the target-local GDF-corrected MSPE criterion at the
target input, with no additional rolling-validation step.

\begin{table}[t]
\centering
\caption{\textbf{Target-relevant variable selection using the target-local
GDF-corrected MSPE criterion.} The candidate set contains two true covariates
\(\{x^{(1)},x^{(2)}\}\) and two pure noise covariates
\(\{z^{(1)},z^{(2)}\}\), giving \(2^4=16\) candidate subsets. The selected
model minimizes the proposed target-local GDF-corrected MSPE criterion. Entries
report Monte Carlo mean out-of-sample MSPE with empirical standard deviation in
parentheses over \(R=300\) replications. The last column reports the percentage
of replications in which the selected subset is exactly
\(\{x^{(1)},x^{(2)}\}\).}
\label{tab:variable_selection}
\small
\setlength{\tabcolsep}{4.5pt}
\begin{tabular}{cc ccc}
\toprule
\(T\) & \(N\)
& Selected model MSPE (SD)
& Full model MSPE (SD)
& Correct input selected \\
\midrule
 30 &  40 & \textbf{4.371 (3.603)} & 5.463 (3.663) & 100\% \\
 30 & 200 & \textbf{3.709 (2.255)} & 4.341 (2.337) & 100\% \\
120 &  40 & \textbf{3.600 (2.482)} & 4.167 (2.591) & 100\% \\
120 & 200 & \textbf{3.238 (1.874)} & 3.661 (1.960) & 100\% \\
\bottomrule
\end{tabular}
\end{table}
\FloatBarrier

Table~\ref{tab:variable_selection} shows that the target-local MSPE criterion
selects the correct input set in all replications across the four panel
settings. The selected model also has lower MSPE than the full model using all
four variables. This indicates that irrelevant variables can harm StaLoP not
only through overfitting, but also by changing distances, local bases,
state-vector construction, panel discrepancy scores, and relevance weights. In
this controlled design, the proposed MSPE criterion removes the irrelevant
variables and improves prediction. These results demonstrate target-relevant
variable recovery in this setting and should not be interpreted as a general
variable-selection consistency result.

%% file: real_data.tex
% Requires: \usepackage{xcolor}

%========================================================

\section{Real Data: County-to-County Migration Flows}
\label{sec:realdata}

We apply the proposed framework to county-to-county origin--destination (OD)
migration flows aggregated in semiannual bins. This application follows a
digital-twin feedback loop: as each new semiannual migration panel arrives, it
is added to the historical system, StaLoP is updated, and the updated model is
used to predict future OD flows using current origin, destination, distance,
and flood-event information. The next observed panel can then be used to
evaluate and update the system again. The goal of the analysis is to assess
whether StaLoP can identify historically relevant migration periods for future
OD-flow prediction when the most recent period is not necessarily the most
informative. The estimated target-local relevance weights also provide a
diagnostic for understanding which historical periods contribute most to each
prediction.

The raw file contains 23,079,936 monthly OD records, covering 3,008 origin
counties and 3,025 destination counties from 2000 to 2019. The monthly records
are aggregated into 40 semiannual bins. After restricting to Georgia-origin
flows and aggregating to semiannual OD edges, the analysis contains 237,950
semiannual aggregated OD records, with 159 Georgia origin counties and 1,616
destination counties. Each semiannual bin contains about 5,949 Georgia-origin
semiannual OD rows on average, with a range of 5,939--5,959, including
zero-flow OD pairs.

Let \(F_{ij,t}\) denote the observed flow count from origin county \(i\) to
destination county \(j\) in bin \(t\). We use the transformed response
\(Y_{ij,t}=\log\{1+\max(F_{ij,t},0)\}\), where the \(\max(\cdot,0)\) operation
handles nonpositive values in the raw data. The analysis uses the full
migration event file with distance and county-level covariates. We focus first
on Georgia-origin OD flows and evaluate rolling one-step-ahead prediction over
the six target bins \(201601\), \(201602\), \(201701\), \(201702\), \(201801\),
and \(201802\), where the bin labels use a year--semester format. For each
target bin \(t_b\), all earlier bins are used for training. Prediction is
evaluated on active historical OD edges, defined as OD pairs with positive flow
in at least one earlier bin; their target-bin flow may still be zero. This
protocol focuses evaluation on OD links with historical migration evidence,
rather than on the larger set of county pairs that are structurally inactive
over the observed history. For the six rolling target bins, the number of
active evaluation edges ranges from 86 to 105, with a mean of 93.5. The
corresponding historical training sets contain between 190,278 and 220,073
Georgia-origin semiannual aggregated OD records.

We use StaLoP to estimate the conditional mean function of \(Y_{ij,t}\) using
inputs that describe the origin, destination, time bin, distance between the
origin and destination counties, and the number of flood events at the
destination during the corresponding time bin. For each target bin \(t_b\), all
earlier bins are used as historical panels. StaLoP constructs one-sided
target-local state vectors from the historical panels, measures their
target-local compatibility with the target bin, and forms predictions by
borrowing more heavily from compatible historical periods. The retained-panel
count is selected by the target-local GDF-corrected MSPE criterion in
Section~\ref{sec:optimal_weight_selection}. Covariates observed at the target
bin are used only as prediction inputs, while all model fitting and
state-vector construction use training data from earlier bins.

We compare four methods under the same prediction protocol. Global Linear pools
historical information without localization in time or state. Time-Local uses a
one-sided time-local smoother. KNN-Covariate uses covariate localization for
the origin and destination panel effects. StaLoP combines target-local
predictive compatibility with covariate localization and selects the
retained-panel count by the target-local GDF-corrected MSPE criterion. This
comparison separates global pooling, temporal proximity, covariate
localization, and target-local borrowing.

\begin{table}[t]
\centering
\caption{\textbf{Rolling out-of-sample MSPE for Georgia-origin active OD
edges.} Results are reported over semiannual target bins from 2016 to 2018.
Active edges are OD pairs with positive flow in at least one previous bin.
StaLoP uses the target-local GDF-corrected MSPE criterion to select the
retained-panel count. Lower MSPE is better.}
\label{tab:realdata_current_methods_2016_2018}
\small
\setlength{\tabcolsep}{5.5pt}
\begin{tabular}{c c c c c c}
\toprule
Target bin
& \(n_{\mathrm{eval}}\)
& Global Linear
& Time-Local
& KNN-Covariate
& StaLoP \\
\midrule
201601 &  86 & 0.0694 & 0.1554 & 0.0688 & \textbf{0.0285} \\
201602 &  90 & 0.0736 & 0.2181 & 0.0729 & \textbf{0.0671} \\
201701 &  90 & 0.0749 & 0.0793 & 0.0742 & \textbf{0.0627} \\
201702 &  93 & 0.0891 & 0.1310 & \textbf{0.0887} & 0.0920 \\
201801 &  97 & 0.0609 & 0.2679 & \textbf{0.0603} & 0.0720 \\
201802 & 105 & 0.0642 & 0.2470 & \textbf{0.0637} & 0.0685 \\
\midrule
Mean & -- & 0.0720 & 0.1831 & 0.0714 & \textbf{0.0651} \\
\bottomrule
\end{tabular}
\end{table}

Table~\ref{tab:realdata_current_methods_2016_2018} reports the Georgia-origin
rolling prediction results. StaLoP has the lowest average MSPE, with mean MSPE
\(0.0651\), compared with \(0.0720\) for Global Linear, \(0.1831\) for
Time-Local, and \(0.0714\) for KNN-Covariate. The largest improvement is
relative to Time-Local, showing that the most recent bins are not always the
best source of information for predicting migration flows. StaLoP also improves
on Global Linear and KNN-Covariate on average, although the bin-level
comparison is heterogeneous. StaLoP has the lowest MSPE in the first three
target bins. In the last three bins, KNN-Covariate is slightly lower, with
differences of \(0.0034\), \(0.0117\), and \(0.0048\) in MSPE for 201702,
201801, and 201802, respectively. These results suggest that target-local
borrowing improves average prediction, while covariate-local structure remains
competitive in some periods.

\begin{table}[t]
\centering
\caption{\textbf{Robustness across high-volume origin states.}
Entries report average rolling out-of-sample MSPE over semiannual active-edge
target bins from 2016 to 2018. Active edges are OD pairs with positive flow in
at least one earlier bin. Lower MSPE is better.}
\label{tab:realdata_multistate_robustness}
\small
\setlength{\tabcolsep}{5.5pt}
\begin{tabular}{lccccc}
\toprule
Origin state
& Mean \(n_{\mathrm{eval}}\)
& Global Linear
& Time-Local
& KNN-Covariate
& StaLoP \\
\midrule
Georgia &  93.5 & 0.0720 & 0.1831 & 0.0714 & \textbf{0.0651} \\
Florida & 228.2 & 0.5413 & 0.8189 & 0.5357 & \textbf{0.5101} \\
Texas   & 195.8 & 0.9427 & 1.0700 & 0.9381 & \textbf{0.8985} \\
\bottomrule
\end{tabular}
\end{table}

Table~\ref{tab:realdata_multistate_robustness} reports the same active-edge
evaluation for Georgia, Florida, and Texas. StaLoP achieves the lowest average
MSPE in all three states. The improvement is again largest relative to
Time-Local. The comparison with Global Linear and KNN-Covariate is smaller but
remains favorable on average. Thus, the predictive advantage of target-local
borrowing is not specific to the Georgia detailed analysis. At the same time,
the state-level averages should not be interpreted as uniform dominance in
every target period, as the Georgia bin-level results show that the best method
can vary across bins.

\begin{figure}[t]
\centering
\includegraphics[width=\textwidth]{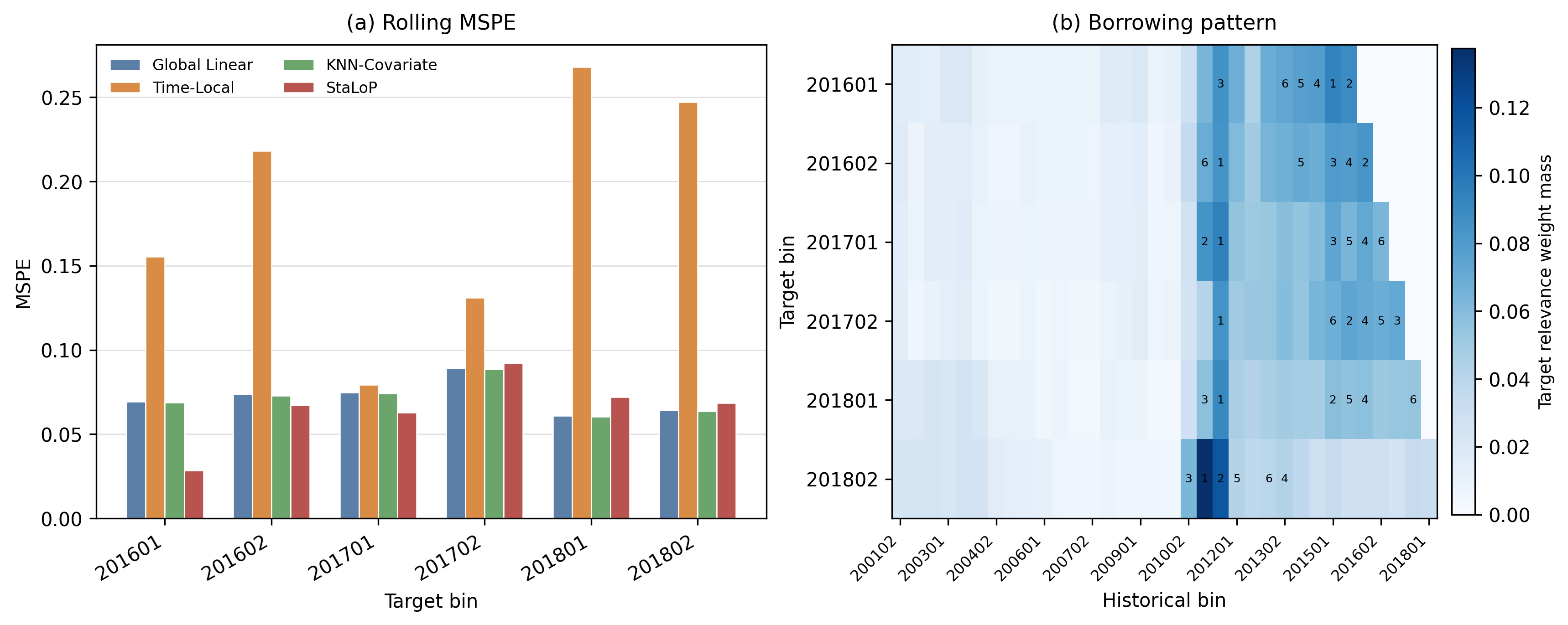}
\caption{\textbf{Real-data county-to-county OD migration-flow prediction and
StaLoP borrowing pattern.}
Panel (a) reports rolling out-of-sample MSPE for Georgia-origin active OD
edges. Lower MSPE indicates better predictive performance. Panel (b) reports
the combined origin-side and destination-side target relevance weight mass
assigned to historical bins for each target bin. Numbers indicate the rank of
each historical bin by target relevance weight mass, with 1 denoting the bin
receiving the largest weight mass.}
\label{fig:realdata_prediction_retrieval}
\end{figure}

Figure~\ref{fig:realdata_prediction_retrieval} summarizes the Georgia-origin
prediction results and the StaLoP borrowing pattern. Panel (a) shows that
Time-Local has larger errors in several target bins, while StaLoP is more
stable across the rolling evaluation periods. Panel (b) shows that the largest
StaLoP relevance weights are often assigned to nonadjacent historical bins
rather than only to the most recent bin. This explains the gain over
Time-Local: StaLoP can retrieve historical periods with stronger target-local
compatibility even when they are not temporally close to the target period. The
estimated relevance weights also make the borrowing mechanism interpretable by
showing which historical bins contribute most to each prediction. Overall, the real-data analysis shows that StaLoP improves average rolling
prediction accuracy and provides interpretable target-local borrowing weights
that help identify historically relevant migration periods for future OD-flow
prediction.

%% file: discussion.tex
\section{Conclusions and Future Directions}
\label{sec:discussion}
We propose StaLoP, a state-local prediction framework for sequential panel data motivated by digital-twin feedback systems. In such systems, new panels arrive over time, the predictive model is updated, and the updated model is used to forecast future system behavior under current input conditions. StaLoP borrows historical information according to target-local predictive compatibility rather than temporal proximity. For a target input \(\mathbf x^\ast\), StaLoP represents historical panels through target-local state vectors, compares these state vectors with the target panel, and combines the resulting target relevance weights with covariate localization. Theoretical results characterize the conditional bias--variance trade-off, establish a conditional normal approximation after accounting for local approximation and transfer bias, justify simultaneous uncertainty bands, and support the target-local GDF-corrected MSPE criterion for retained-panel selection. Simulation studies show that StaLoP reduces MSPE relative to global, time-local, and covariate-local baselines in settings with recurrent time-specific mean-function patterns. The migration-flow application illustrates the digital-twin update loop and shows that StaLoP improves average rolling prediction accuracy for Georgia, Florida, and Texas, while the estimated relevance weights identify which historical panels contribute most to each prediction.

Several extensions remain for future work. The current theory is conditional on the realized historical design and realized prediction weights, which matches the rolling prediction protocol but does not fully account for all data-adaptive steps. A remaining theoretical question is how to obtain unconditional guarantees
that jointly account for target-local state-vector estimation, empirical
panel discrepancy construction, retained-panel selection, and prediction-stage
noise. The migration analysis focuses on active OD edges and therefore measures the intensive margin of established migration flows; an extension is a two-part model that separately studies edge formation and conditional flow intensity. The StaLoP framework can also be extended to richer state representations, including learned or structured representations, provided that the construction uses only information available before the target response is evaluated. In digital-twin applications, these extensions would allow StaLoP to update
sequentially as new panels arrive and to provide both predictions and
historical panels with high target relevance for decision support.

%\section*{Data Availability Statement}

%The data and code supporting the findings of this study are openly available
%in the GitHub repository
%\url{https://github.com/jasonhanrh/state_local}.